\documentclass{PoS}
\usepackage{makeidx}

\newcommand{\be}{\begin{eqnarray}}
\newcommand{\ee}{\end{eqnarray}}
\newcommand{\rar}{\rightarrow}

\usepackage[]{caption}
\def\mc{\mathcal}
\def\mpl{M_{Pl}}

\def\-g{\sqrt{-g}}
\renewcommand\rho{\varrho}

\usepackage[utf8]{inputenc}
\usepackage{graphicx,array,dcolumn,xcolor}

\title{Instabilities in modified theories of gravity}

\ShortTitle{Instabilities in modified theories of gravity}

\author{\speaker{Elena Arbuzova}\thanks{This work was supported by the RSF Grant 19-42-02004.}\\
        Dubna State University and Novosibirsk State University\\
        E-mail: \email{arbuzova@dubna-uni.ru}}

\abstract{The review is devoted to consideration of possible observational consequences of modified gravity theories, suggested for explanation  of
the contemporary accelerated expansion of the universe. The major attention is paid to $F(R)$-models. It is shown that in systems with rising energy density high frequency and  large amplitude oscillations of the curvature scalar, $R(t)$, are induced. These oscillations lead to the production of elementary particles, which may be observed in the spectra of energetic cosmic rays. In the background of such oscillating solutions gravitation repulsion between finite-size objects becomes possible. Since the Lagrangian is a non-linear function of curvature, equations of  motion become of higher (4th) order and exhibit very rich pattern of new physical effects. In particular, the evolution of density perturbations is strongly different from that 
in General Relativity, amplified due to both parametric resonance and anti-friction phenomena.   
          }

\FullConference{Multifrequency Behaviour of High Energy Cosmic Sources - XIII - MULTIF2019\\
		3-8 June 2019\\
		Palermo, Italy}

\begin{document}

\section{Introduction}

The most impressive events in astronomy made during approximately the last two decades were a discovery and accumulation of independent pieces of evidence proving that our universe expands with acceleration, but the driving force behind this accelerated expansion is still unknown.  

There is a large set of independent data based on completely different cosmological and astrophysical phenomena, confirming that the expansion rate of the universe began to increase with time at the relatively recent cosmological epoch, at redshifts of the order of unity ($ z_ {acc} \approx 0.65 $).

Among the observational data proving the existence of accelerated cosmological expansion, the most impressive are the direct measurements of the acceleration by dimming of distant type Ia supernovae, carried out in 1998. For discovering this amazing effect Sol Perlmutter, Brian Schmidt and Adam Riess received the Nobel Prize in Physics in 2011 \cite{Riess:1998cb,Perlmutter:1997zf,Schmidt:1998ys,Perlmutter:1998np,Riess:2004nr}.

The accelerated cosmological expansion of the universe is supported not only by the data obtained by measuring the luminosity of supernovae, but also by other independent astronomical observations, including the problem of the universe age, which arose in the 1980s. The universe age calculated with known at that time cosmological parameters 
\cite{Freedman:1994tr,Schmidt:1994fu,Riess:1996pa}
turned out to be significantly less than the age of old stars and star clusters
\cite{Chaboyer:1996pa,Chaboyer:1995zs,Chaboyer:1995zn}.  

The position of the first peak in the spectrum of angular fluctuations of cosmic microwave background (CMB) 
\cite{Hinshaw:2012aka,Aghanim:2018eyx} 
indicates that the geometry of the universe is very close to flat, Euclidean. To implement this the matter density in the universe should be close to the critical energy density. However, the direct measurements of matter density 
\cite{Percival:2001hw,Peacock:2001gs}
give three times smaller value, which means an existence of some kind of new form of matter in the universe.

An analysis of large scale structure of the universe \cite{Deustua:2002yx,Coil:2012vw} also indicates that, in addition to the matter creating usual gravitational attraction, there must exist some unknown substance called {\it dark energy} that exhibits antigravitational properties. It suppresses the cosmic structure formation at very large scales and has impact on CMB angular fluctuations  spectrum. 

The universe expansion is described by the Hubble parameter, $H=\dot a/a$, where $a(t)$ is a scale factor, which satisfies the Friedmann equations. 
The cosmological acceleration is determined by the second Friedmann equation:
\be
\frac{\ddot a}{a} = -\frac{4\pi\,G_N }{3}\,(\rho+3P), 
\ee 
where $G_N$ is a gravitational constant, $\rho = \rho(t)$ is the energy density of matter and $P$ is a pressure. 

As we see from the presented equation the source of cosmological acceleration is not only the energy density as we would expect in the Newtonian theory, but  pressure also gravitates. With negative pressure the cosmological acceleration becomes positive, if $P<- \rho/3$. Obviously, that negative energy density also may lead to cosmic antigravity, but the theories with $\rho<0$ are pathological and are usually not considered. 

It should be emphasized that antigravity in General Relativity (GR) is possible only for infinitely large objects. Any object of finite size with a positive energy density can create only gravitational attraction, as states the Jebsen-Birkhoff theorem, well-known in GR \cite{Jebsen,Birk,Hawking:1973uf}. However, as it is shown in Sec.~\ref{sec-antygrav}, in $F(R)$-modified gravity this theorem is violated 
and finite size objects may cause  gravitational repulsion.  

Conclusions about the presence in the universe of an unusual form of matter/energy,  {\it dark energy}, were made in the framework of Einstein's theory of gravity, General Relativity. An alternative explanation of the aforementioned astronomical data in favor of cosmological acceleration can be obtained on the basis of gravity modification at cosmologically large distances. Both approaches explain the accelerated expansion of the universe, since they lead to the effective replacement of gravitational attraction by gravitational repulsion on cosmological scales. The difference between them is that dark energy is a source of antigravity, while the modification of gravity leads to a change of the sign of gravitational force even in the absence of matter.

In modified $f(R)$-theories of gravity the usual Einstein-Hilbert action acquires an additional term, non-linear $F(R)$-function, which changes gravity at large distances and is responsible for cosmological acceleration:
\be 
S = \frac{M_{Pl}^2}{16\pi} \int d^4 x \sqrt{-g} [R+  F(R)]+S_m\,,
\label{grav-mdf}
\ee
where $M_{Pl}= 1. 2 2\cdot 10^{19}$ GeV is the Planck mass and $S_m$ is the matter action. 

The usual General Relativity action is linear in the curvature scalar $R$. This is the reason why the GR 
equations contain, as it is usual in other field theories, only second derivatives of metric despite the
fact that the action also contains second derivatives. If the action differs from a simple linear GR form, the equation of motion would  be higher than the second
order one. Such equations should contain some pathological features as existence of tachyonic solutions
or ghosts. However, the theories whose action depends only on a function of the curvature scalar, $F(R)$,
are free of such pathologies because, as is known, they are equivalent to an addition of a scalar degree of 
freedom to the usual GR with the scalar field satisfying normal second order field equation. That's why
modifications of gravity at large distances are mostly confined to  ${ F(R) }$-theories, though the conditions of stability and/or of the absence of singularities impose certain restrictions on the form of $F(R)$-function. 

The function $F(R)$ is chosen is such a way that the gravitational equations of motion which replace the usual Einstein equations have an accelerated de Sitter-like solution with a constant curvature, $R$, even in the absence of matter. 

The pioneering works in this direction were done in Ref.~\cite{Capozziello:2002rd,Capozziello:2003gx,Capozziello:2003tk}, which were closely followed by 
Ref.~\cite{Carroll:2003wy,Carroll:2004de}. In these works the singular in $R$ action
\be
F(R) = - {\mu^4}/{R} \,,
\label{F-of-R}
\ee
has been explored. Constant parameter $\mu$ was chosen as
${\mu^2 \sim R_c \sim 1/t_U^2}$ to  describe the observed cosmological acceleration, where $R_c$ is a present day average curvature of the universe, $t_U=4\cdot10^{17}$~sec is a universe age. 

However, as it was shown by A. Dolgov and M. Kawasaki~\cite{ad-mk-instab} such a choice $F(R)$-function leads to a strong instability in the presence of matter due to  the small coefficient, ${ \mu^4}$, in front of the highest derivative in the corresponding equations of motion. Small fluctuations grow exponentially with the characteristic time  
\be
\tau=\frac{\sqrt{6}\mu^2} {T^{3/2} }
\sim 10^{-26} {\rm sec}  
\left(\frac{\rho_m}{{\rm g/ cm}^{3} }\right)^{-3/2}\, ,
\label{t-eff}
\ee
where ${T=8\pi T_\mu^\mu/M_{Pl}^2 \sim (10^{3} {{ sec}})^{-2} \left({\rho_m}/{{ g/ {cm}}^{-3}}\right) }$ with $T_\mu^\mu$ being the trace of the energy-momentum tensor of matter,  ${\rho_m}$ is the mass density of the celestian  body. 

To avoid the problem of such instability modified gravity was further modified. The class of the models, which are free from exponential instability, is described in  review ~\cite{appl-bat-star}. Various functions presented in the works~\cite{Hu:2007nk,Appleby:2007vb,Starobinsky:2007hu}, have the form:
\be \label{eq:cr1}
F_{ HS}(R) & = &  - \frac{R_{ vac}}{ 2} \frac{c \left(\frac{R }{ R_{ vac}}\right)^{2n}}
 { 1+ c \left(\frac{R }{ R_{ vac}}\right)^{2n}}\,, \\   \label{eq:cr201} 
F_{AB}(R) & = &   \frac{\epsilon}{ 2}\,
\log \left[ \frac{\cosh\left(\frac{R }{ \epsilon}-b\right) }{ \cosh b} \right]  
 - \frac{R}{2}\,, \\
F_S(R) &=& \lambda R_c \,\left[ {\left(1+ \frac{R^2}{R_c^2}\right)^{-n}} - 1 \right]\,.
\label{F-AAS}
\ee
These functions were carefully constructed to satisfy a number of conditions to avoid pathologies and ensure the stability of cosmological solutions in the future, as well as classical and quantum stability (gravitational attraction and absence of ghosts). Despite different forms they result in quite similar consequences. 
Some other forms of gravity modification can be found in the review~\cite{odin-rev}.

Despite considerable improvement, the models (\ref{eq:cr1}) - (\ref{F-AAS}) possess
another troublesome feature, namely in a cosmological situation they should evolve from a singular 
state with an infinite ${R}$ in the past~\cite{appl-bat-08}. In other
words, if we travel backward in time from a normal cosmological state, we come to a singular state with 
infinite curvature while the energy density remains finite.

In cosmology energy density drops down with time and singularity doesn't appear in the future. However,  systems with rising mass/energy density will evolve to a
singularity, $R \rar \infty$,  in a finite time~\cite{frolov,Thong-Sami-1,Thong-Sami-2,Arbuzova:2010iu,Reverberi:2012ew}. Such future singularity is unavoidable, regardless of the initial conditions, and infinite value of $R$ would be reached in time which is much shorter than the cosmological time.

\section{Instability and curvature oscillations in $F(R)$ theories \label{inst-osc}}

Following ref.~\cite{Arbuzova:2010iu} let us consider version (\ref{F-AAS}) of $F(R)$ function in the case of large $R$.
We analyze the evolution of $R$ in massive objects with time  varying mass density,
$\rho_m \gg \rho_c $. The
cosmological energy density at the present time is 
$\rho_c \approx 10^{-29}\,{\rm g/cm}^3$, while matter density of, say,
a dust cloud in a galaxy could be about $\rho_m \sim 10^{-24}{\rm g/cm}^3$. Since the magnitude of the curvature scalar is proportional to the
mass density of a nonrelativistic system, we find $R \gg R_c $. In this limit we can approximately take:
\be 
F(R) \approx -\lambda R_c \left[ 1 -\left(\frac{R_c}{R}\right)^{2n} \right] \,.
\label{F-large-R}
\ee

The equations of motion which follow from the action (\ref{grav-mdf}) have the form:
\be
\left( 1 + F'_{R}\right) R_{\mu\nu} -\frac{1}{2}\left( R + F\right)g_{\mu\nu}
+ \left( g_{\mu\nu} D_\alpha D^\alpha - D_\mu D_\nu \right) F'_{R} = 
\frac{8\pi T^{(m)}_{\mu\nu}}{M_{Pl}^2}\,,
\label{eq-of-mot}
\ee 
where $F'_{R}= dF/dR$, $D_\mu$ is the covariant derivative, and $T^{(m)}_{\mu\nu}$ is the energy-momentum tensor of matter. Taking the trace of Eq.~(\ref{eq-of-mot}) leads to the equation:
\be
3 D^2 F'_R -R + R F'_R - 2F = 8\pi T_\mu^\mu /M_{Pl}^2\,.
\label{trace-0}
\ee
This is a closed equation for $R$ except for metric depending terms in the covariant derivative and in $T^\mu_\mu$. If the metric slightly 
differs from the flat Minkowski
one, equation (\ref{trace-0}) would contain only one unknown scalar function which completely determines the evolution of $R$.
In this limit the equation can be reduced to the simple oscillator form:
\be
(\partial^2_t - \Delta) w + U'(w) = 0
\label{eq-for-w}
\ee
for the function
 \be
w \equiv - F'_R \approx 2n\lambda \left(\frac{R_c}{R}\right)^{2n+1} \,,
\label{F'-of-R}
\ee
where the potential $U(w)$ is defined as:
\be
U(w) = \frac{1}{3}\left( \tilde T - 2\lambda R_c\right) w + 
\frac{R_c}{3} \left[ \frac{q^\nu}{2n\nu} w^{2n\nu}+ \left(q^\nu
+\frac{2\lambda}{q^{2n\nu} } \right) \,\frac{w^{1+2n\nu}}{1+2n\nu}\right]\,,
\label{U-of-w}
\ee
with $\nu = 1/(2n+1)$, $q= 2n\lambda$, $U'(w)=dU/dw$, and $\tilde T = 8\pi T_\mu^\mu /M_{Pl}^2$. Moreover, we are considering the case $|R| \gg R_c$, since in realistic astrophysical systems $\tilde T\gg R_c$. Their ratio is about $\tilde T/R_c \sim \rho_m/\rho_c \gg 1$ and hence $w\ll 1$. Thus the first term in square brackets in Eq.~(\ref{U-of-w}) dominates. The potential $U$ would depend on time if the mass
density of the object under scrutiny also changes with time, $\tilde T=\tilde T(t)$.

If only the dominant terms are kept in equations (\ref{eq-for-w}) and (\ref{U-of-w}) and if the space derivatives are neglected, the equation (\ref{eq-for-w}) simplifies to:
\be
\ddot w + \tilde T/3 - \frac{q^\nu (-R_c)}{3w^\nu}=0\,.
\label{eq-w-simple}
\ee 
It is convenient to introduce the dimensionless quantities:
\be
t = \gamma \tau,\,\,\, w = \beta \zeta\,,
\label{dimless}
\ee
where $\beta$ and $\gamma$ are so chosen (see below) that the equation for {$\zeta$} becomes particularly simple:
\be
\zeta'' - \zeta^{-\nu} + (1+\kappa \tau) = 0\,.
\label{eq-for-z}
\ee
Here a prime denotes differentiation with respect to $\tau$ and the trace
of the energy-momentum tensor of matter is parametrised as:
\be
\tilde T(t) = \tilde T_0 (1 + \kappa \tau)\,. 
\label{T-of-t}
\ee
The constants $\gamma$ and $\beta$ are equal to
\be
\gamma^2 = \frac{3q}{(-R_c)} \left(-\frac{R_c}{\tilde T_0}\right)^{2(n+1)},\ \ \\
\beta = \gamma^2 \tilde T_0/3 = q \left(-\frac{R_c}{\tilde T_0}\right)^{2n+1}\,.
\label{gamma-beta}
\ee

\begin{figure}[ht]
\begin{center}
\includegraphics[width=.32\textwidth]{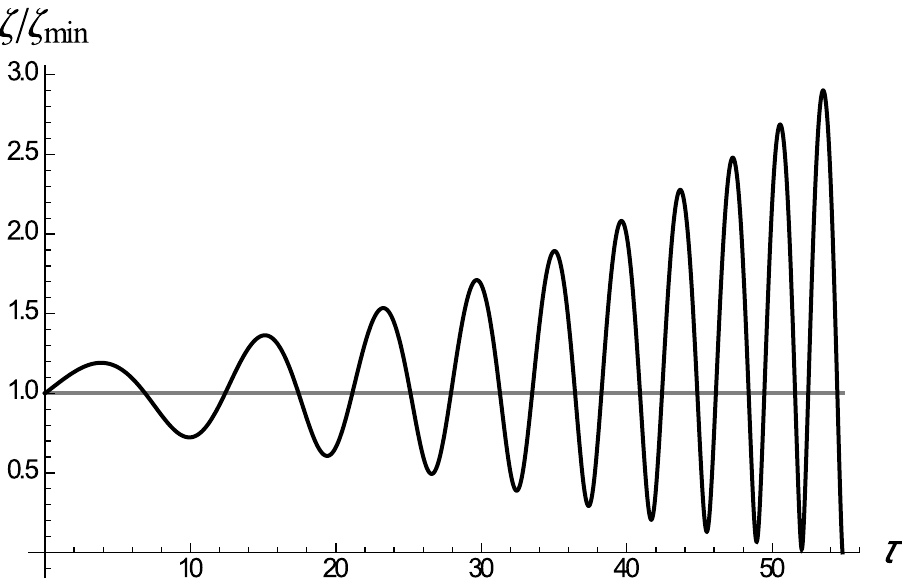} 
\includegraphics[width=.32\textwidth]{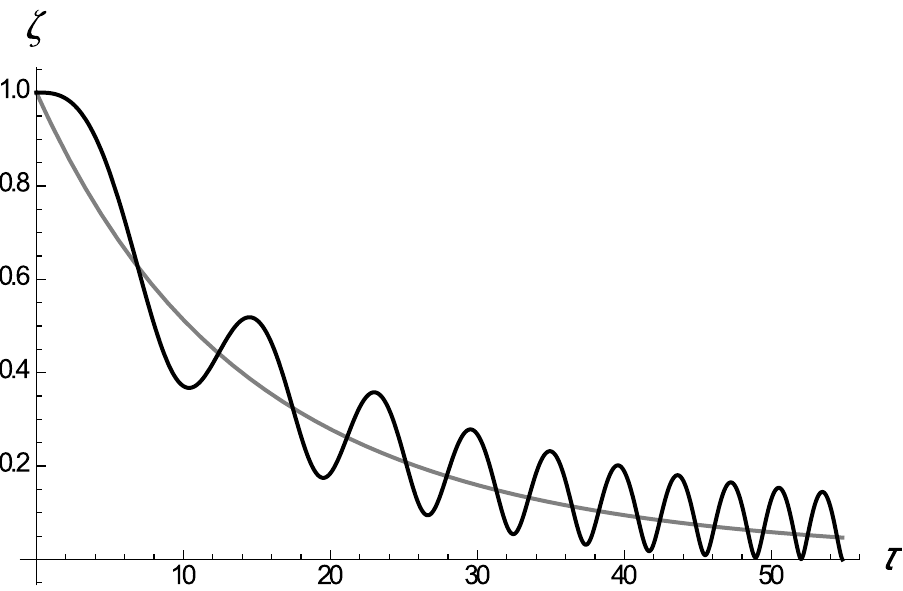} 
\includegraphics[width=.32\textwidth]{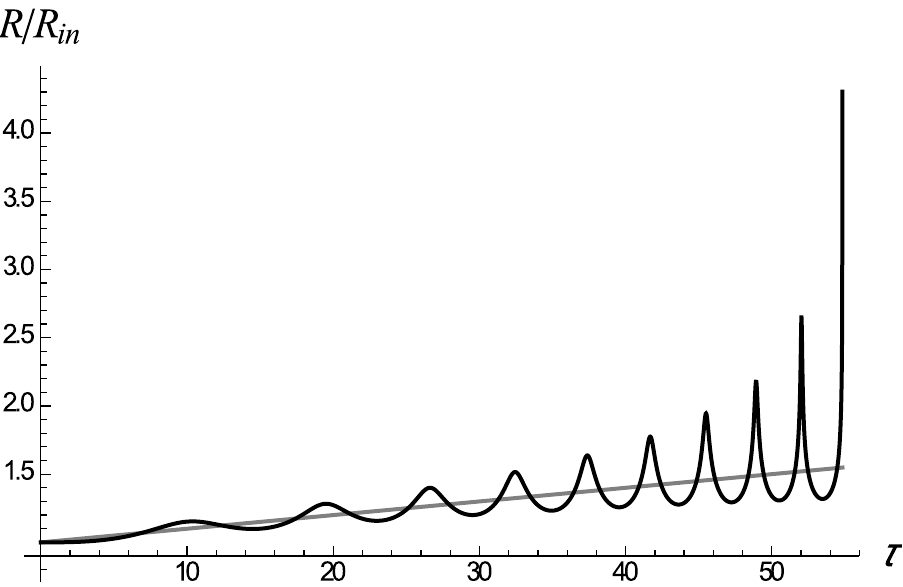}
 \caption{
Ratio $\zeta/\zeta_{min}$ (left), functions $\zeta$ and $\zeta_{min}$ (centre), and corresponding $R$ (right) for
$n=3$, $\kappa = 0.01$, $\rho_m/\rho_c = 10^5$.}
\label{f-34}
\end{center}
\end{figure}

It was shown in Ref.~\cite{Arbuzova:2010iu} that in systems with rising mass density the position of the minimum of the potential $\zeta_{min} = (1+\kappa \tau )^{-1/\nu}$ moves towards zero and so does $\zeta (\tau)$ itself, practically independently on the initial conditions, i.e. on $\zeta(0)$ and $\zeta'(0)$. The function $\zeta (\tau)$ oscillates around $\zeta_{min} (\tau)$ and at some moment it passes beyond the minimum of $U$ and reaches the point $\zeta = 0$. This value of $\zeta$ however corresponds to the 
{singularity} $R\to \infty$. The system arrives to the singularity in a finite time, while the external energy density still remains finite. This conclusion is supported by the numerical calculations, shown in Fig.~
\ref{f-34}. 
For these particular figures we took $\zeta_0 = 1$ (i.e. the GR value) and $\zeta'_0= 0$. 

To eliminate the past and future singularities the function (\ref{F-AAS}) was modified by adding a term proportional to the curvature squared~\cite{Starobinsky:2007hu}:
\be\label{eq:model}
F(R) = -\lambda R_c\left[1-\left(1+\frac{R^2}{R_c^2}\right)^{-n}\right]-\frac{R^2}{6m^2}\,,
\ee
where the parameter
 $m$ has 
{ dimension of energy.} 
The additional term is relevant only at very large curvatures, because $m\gtrsim 10^5$ GeV is necessary in order to preserve 
the successful predictions of the standard BBN~\cite{Arbuzova:2011fu}. 

At small curvatures the system tends to evolve to higher values of $R$, but, as $|R|$ grows, the $R^2$-term becomes dominant and pushes the system to lower values of curvature. This results in oscillating solutions $R(t)$, possibly with very large amplitude.  

Below, follow our works~\cite{Arbuzova:2012su,Arbuzova:2013ina}, we consider the model based on Eq.~(\ref{eq:model}). 
A large $m$ implies that the stabilisation takes place at very high $R$. Though $R$ does not become infinite, it can reach huge values in systems with rising mass density. This rise normally originates after the onset of structure formation at $z\sim 10^4$ or at any time later. 

 We are particularly interested in the regime $|R_c|\ll|R|\ll m^2$, in which $F(R)$ can be approximated by
\be\label{eq:model_approx}
F(R)\simeq -R_c\left[1-\left(\frac{R_c}{R}\right)^{2n}\right]-\frac{R^2}{6m^2}\,.
\ee
We consider a nearly-homogeneous distribution of pressureless matter, with energy/mass density rising with time but still relatively low (e.g. a gas cloud in the process of galaxy or star formation). In such a case the space derivatives can be neglected and, if the object is far from forming a black hole, the space-time metric is approximately Minkowski. Then Eq.~(\ref{trace-0}) takes the form
\be
\label{eq:trace_approx}
3\partial_t^2 F'_{R} - R - \tilde T = 0\,.
\ee
Let us introduce the dimensionless quantities\footnote{The parameter $g$ should not be confused with $\det g_{\mu\nu}$.}
\be\label{eq:definitions}
 z&\equiv& \frac{T(t)}{T(t_{in})}\equiv \frac{\widetilde T}{T_0}= \frac{\varrho_m(t)}{\varrho_{m0}}\,, \ \ \ \ \ \ 
 \qquad y\equiv -\frac{R}{T_0}\,, \\
g&\equiv & \frac{T_0^{2n+2}}{6 n(-R_c)^{2n+1}m^2}= \frac{1}{6 n  (m t_U)^2} \,\left( \frac{\varrho_{m0}}{\varrho_c}\right)^{2n+2}\,,
\qquad \tau\equiv m\sqrt g\,t\,,
\ee
where $\varrho_c \approx 10^{-29} $ g/cm$^3$ is the cosmological energy density at the present time,
$\varrho_{m0}$ is the initial value of the mass/energy density of the object under scrutiny,
and $T_0 = 8\pi \varrho_{m0}/m_{Pl}^2$. Next let us introduce the new scalar field:
\be\label{eq:xi_definition}
\xi\equiv  \frac{1}{2 n}\left(\frac{T_0}{R_c}\right)^{2n+1}F_{,R}  = \frac{1}{y^{2n+1}}-gy\, ,
\ee
in terms of which eq.~(\ref{eq:trace_approx}) can be rewritten in the simple oscillator form:
\be\label{eq:xi_evol}
\xi''+z-y=0\,,
\ee
where a prime denotes derivative with respect to $\tau$. The potential of the oscillator is defined by:
\be
\frac{\partial U}{\partial \xi}= z - y(\xi).
\label{U-prime}
\ee

The minimum of potential $U(\xi)$ is located at $y(\xi) = z(\tau)$, so it moves with time according to 
\be
\xi_{min} (\tau) ={z(\tau)^{-(2n+1)}}-gz(\tau). 
\label{xi-min}
\ee
It is intuitively clear that even if initially $\xi$ takes its GR value $\xi = \xi_{min}$ it would not catch the motion of the minimum
and as a result  $\xi$ starts to oscillate around it. Dimensionless frequency of small oscillations, $\Omega$, is determined by:
\be
\label{eq:frequency_U}
\Omega^2 =  \left. \frac{\partial^2 U}{\partial \xi^2}\right|_{y=z} = \left(\frac{2n+1}{z^{2n+2}}+g\right)^{-1}\,.
\ee
Note that physical frequency is $\omega = \Omega\,m \sqrt g$.

\begin{figure}
\centering
 \includegraphics[width=.45\textwidth]{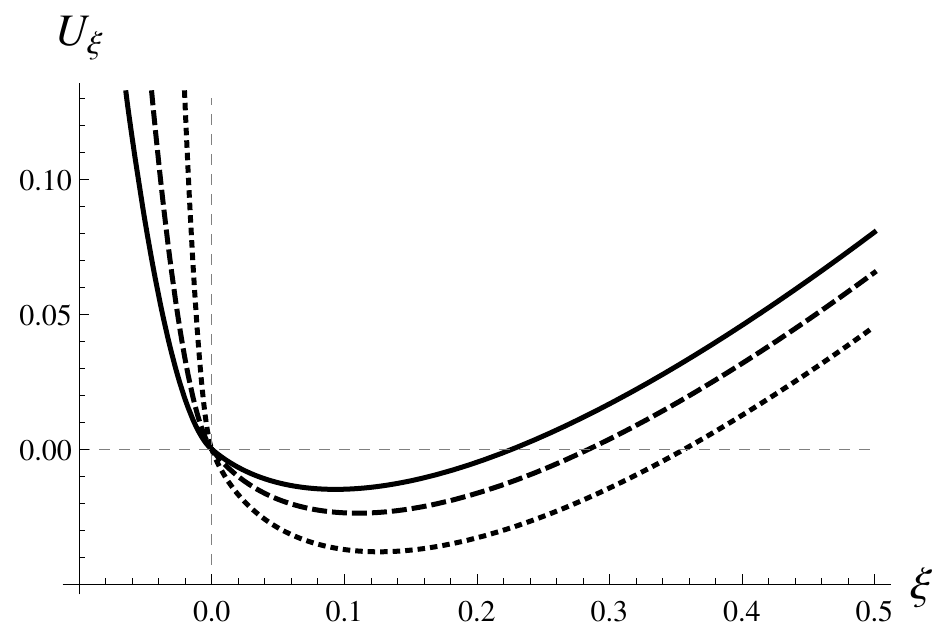}
\includegraphics[width=.45\textwidth]{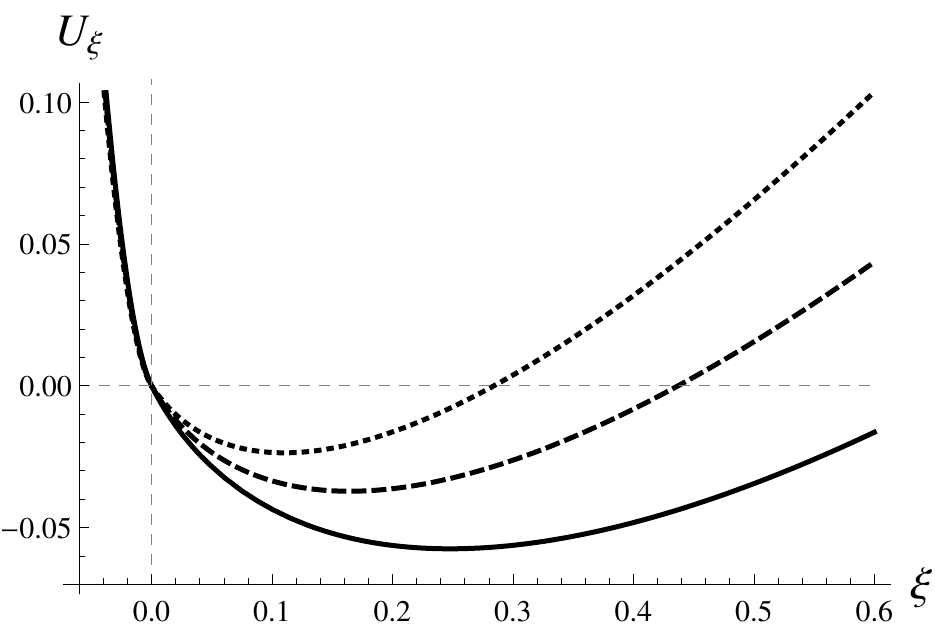}
\caption{Examples of the variation of potential $U(\xi )$ for different values of parameters. 
{\it Left panel} $(n=2,\,z=1.5)$ solid line: $g=0.02$, dashed line: $g=0.01$, dotted line: $g=0.002$.
{\it Right panel} $(n=2,\,g=0.01)$ solid line: $z=1.3$, 
dashed line: $z=1.4$, dotted line: $z=1.5$. The bottom of the potential moves to higher values of $U$ and lower values of $\xi$, 
as $g$ and $z$ increase.}
\label{fig:potential}
\end{figure}

One cannot analytically invert eq.~(\ref{eq:xi_definition}) to find the exact expression for $U(\xi)$. However, we can find approximate expressions for $gy^{2n+2}\ll 1$ ($\xi>0$) and $gy^{2n+2}\gg 1$ ($\xi<0$). The value $\xi=0$ separates two very distinct regimes, in each of which $\Omega$ 
has a very simple expression and $\xi$ is dominated by either one of the two terms in the r.h.s. 
of eq. (\ref{eq:xi_definition}). Hence, in those limits the relation $\xi=\xi(y)$ can be inverted giving an explicit expression for $y=y(\xi)$, and therefore the following form for the potential:
\be\label{eq:xi_pot_theta}
U(\xi) = U_+(\xi)\Theta(\xi) + U_-(\xi)\Theta(-\xi)\,,
\ee
where
\be\label{eq:potential_+_-}
U_+(\xi) &= &z\xi - \frac{2n+1}{2n}\left[\left(\xi+g^{(2n+1)/(2n+2)}\right)^{2n/(2n+1)}-g^{2n/(2n+2)}\right]\,,\\
U_-(\xi) &= & \left(z-g^{-1/(2n+2)}\right)\xi+\frac{\xi^2}{2g}\,.
\ee
By construction $U$ and $\partial U/\partial\xi$ are continuous at $\xi=0$. The shape of this potential is shown in Fig.~\ref{fig:potential}.

The bottom of the potential, as it is obvious from Eq.~(\ref{U-prime}), corresponds to the GR solution $R=- \tilde T$, or $y(\xi) = z$, 
and its depth (for $gz^{2n+2}<1$) is
\be\label{eq:potential_bottom}
U_0(\tau) \simeq -\frac{1}{2n\,z(\tau)^{2n}}\,.
\ee

\begin{figure}[!t]
\centering
 \includegraphics[width=.40\textwidth]{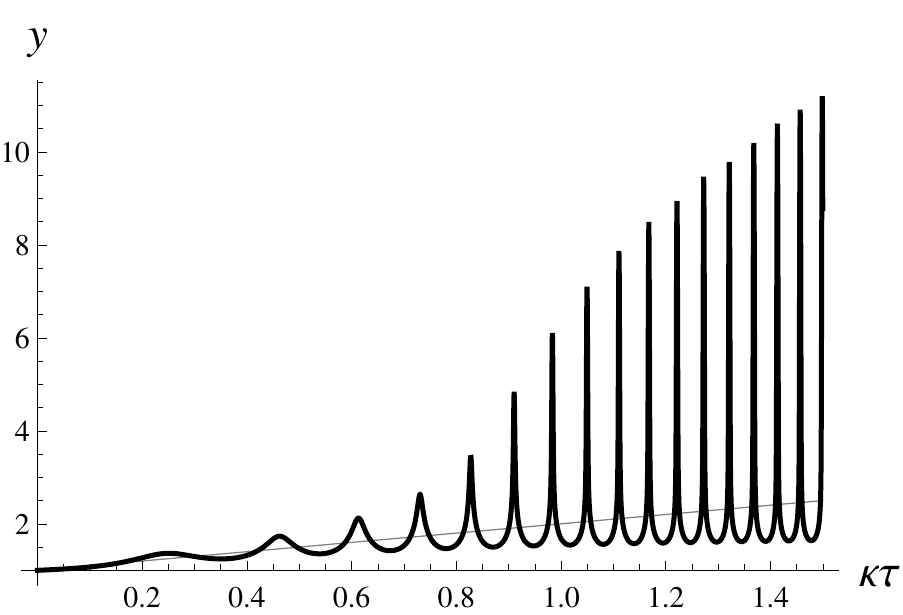} \ \ \ 
\includegraphics[width=.40\textwidth]{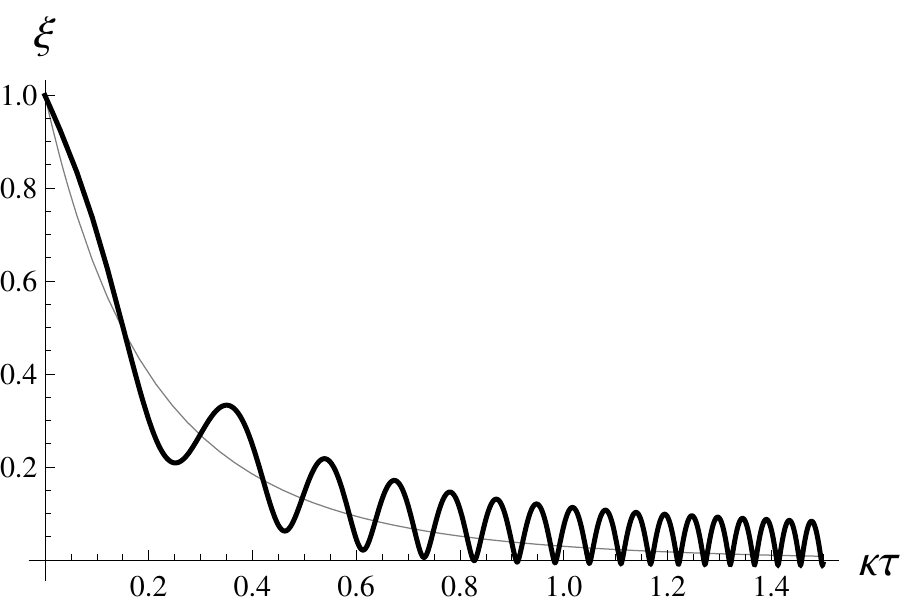}
\caption{``Spikes'' in the solutions. The results presented are for $n=2$, $g=0.001$, $\kappa=0.04$, and $y'_0=\kappa/2$. 
Note the asymmetry of the oscillations of $y$ around $y=z$ and their anharmonicity.}
\label{fig:spikes}
\end{figure}

Our primary goal is to determine the amplitude and shape of the oscillations of $y$. Let us stress that, in contrast to $\xi$, the oscillations of $y$ quickly become strongly anharmonic and even for slightly negative $\xi $ the amplitude of $y$ may be very large because $y \approx -\xi /g$, according to Eq.~(\ref{eq:xi_definition}). This feature is well demonstrated by the results of numerical calculations shown in Fig.~\ref{fig:spikes}. 

According to calculations of  ref.~\cite{Arbuzova:2011fu}, harmonic oscillations of curvature with frequency $\omega$
and amplitude $R_{max}$ transfer energy to massless particles with the rate (per unit time and volume):
\be
\label{eq:PP_harmonic}
\dot\rho_{PP}\simeq {R^2_{max}\,\omega}/{(1152\pi)}\,,
\ee
The life-time of such oscillations is $\tau_R = {48\,\mpl^2}/{\omega^3} $.

In our case the oscillations are far from harmonic and we have to make Fourier expansion of the spiky function $y(\tau)$.
To this end we approximate the ``spike-like'' solution as a sum of gaussians with amplitude modulated by slowly varying
amplitude ${ B(t)}$, superimposed  on smooth background $ A(t)$:
\be
\label{eq:gaussians}
R(t) = A(t)+B(t)\sum_{j=1}^{N}\exp\left[-\frac{(t-jt_1)^2}{2\sigma^2}\right]\,.
\ee
We  assume that $ \sigma\ll t_1 $, { that is} the spacing between spikes is much larger than their width. 
The Fourier transform of { expression} (\ref{eq:gaussians}) is straightforward but rather tedious (details can be
found in our work~\cite{Arbuzova:2013ina}).
Finally we find:
\be
\left|\tilde{\mc R}(\omega)\right|^2\simeq \frac{4\pi^2 B^2
\sigma^2e^{-\omega^2\sigma^2}\Delta t}{t_1^2}\sum_j\delta\left(\omega-\frac{2\pi j}{t_1}\right)\,.
\ee
Identifying $B$ with $R_{max} = y_{max}  T_0$ and integrating over frequencies we obtain
\be
 \dot\rho_{PP} \simeq 
{y_{max}^2 T_0^2}/({576\pi\,t_1})\,.
\label{dot-rho-total}
\ee
Time interval $t_1$ is approximately equal to $2\pi/\omega_{slow}= 2\pi/(\Omega_{slow}m \sqrt{g})$, 
where $\Omega_{slow}$ is given by eq.~(\ref{eq:frequency_U}). 
Taking all the factors together we finally obtain:  
\be
\dot \rho_{PP} = C_n \frac{\rho_0^2 (m t_U)^2\, z^{n+1}}{M_{Pl}^4 t_{contr}}\left( \frac{t_U}{t_{contr} }\right)^{\frac{n-1}{3n+1}}\,
\left( \frac{\rho_c}{\rho_0} \right)^{\frac{(n+1)(7n+1)}{3n +1}},
\label{dot-rho-fin}
\ee
where
\be
C_n = (2n+1)^{\frac{9n-1}{2(3n+1)}} \left( 6\lambda n \right)^{\frac{7n+1}{2(3n +1)}}/ (18 n).
\label{c-n}
\ee

It is convenient to present numerical values: $\rho_c/ m_{Pl}^4 \approx 2\cdot 10^{-123}$ and $ (mt_U)^2 \approx 3.6\cdot 10^{93} m_5^2$, where
$\rho_c \approx 10^{-29}$g/cm$^3$ and  $m_5 = m/10^5$ GeV. Now assuming that the particle production lasts during time $t \approx t_{contr}$
and taking $\rho_0 = \rho_c$, we find the integrated over energy flux of cosmic rays produced by oscillating curvature:
\be
\rho_{CR} \approx 10^{-24} m_5^2 \, z^{n+1} \, {\rm GeV\, s^{-1}\, cm^{-2}}.
\label{rho-CR}
\ee
This result is a lower limit of the flux of the produced particles. With larger $z$ when the minimum of the potential 
shifts deep into the negative $\xi$ region the production probability significantly rises.

For $m =  10^{13}$ GeV the predicted flux of cosmic rays with energy around $10^{20}$ eV could explain the observed anomaly in the spectrum of 
ultra-high energy cosmic rays.

\section{Spherically symmetric solutions in modified gravity and gravitational repulsion \label{sec-antygrav} } 

A detailed study of the solutions of the modified gravity equations in the present day universe
was performed in ref.~\cite{Arbuzova:2012su,Arbuzova:2013ina} for finite-size astronomical objects. 
It was found that if the energy density rises with time, fast oscillations of the scalar curvature are induced, with an amplitude possibly much larger than the usual General Relativity value $R=-\tilde T$. The solution has the form:
\be
R = R_{GR} (r) y (t),
\label{R-of-t}
\ee
where $R_{GR} = - \tilde T(r)$ is the would-be solution in the limit of GR, while the quickly oscillating function
$y(t)$ may be much larger than unity. According to ref.~\cite{Arbuzova:2013ina} the maximum value of $y$ in the so-called spike region is:
\be
y(t) \sim 6 n (2n+1)  m t_U \left(\frac{t_U}{t_{contr} }\right) \left[\frac{\rho_m (t)}{\rho_{m0}} \right]^{(n+1)/2}
\left(\frac{\rho_c}{\rho_{m0}} \right)^{2n+2} ,
\label{y-of-t}
\ee 
where $t_U$ is the universe age, $t_{contr} $ is the characteristic contraction time, so the energy density of the contracting cloud behaves as $\rho_m (t) = \rho_{m0} (1 + t/t_{contr})$, with $\rho_{m0} $ being the initial energy density of the cloud, and $\rho_c~=~10^{-29}$~g/cm$^3$ being the present day cosmological energy density. 
As it was mentioned in the previous section, 
the mass parameter $m$ entering eq.~(\ref{eq:model}) should be larger than about $10^5$ GeV to avoid a conflict with BBN. So the factor $m t_U $ is huge: $ m t_U \geq 10^{47}$ and $y$ can  reach a very high value, if not suppressed by a small ratio 
$( \rho_c/\rho_{m0})^{2n+2}$, when $n$ is  large.

As shown in ref.~\cite{Arbuzova:2013ina}, such high amplitude spikes are formed if 
\be
6 n^2 (2n+1)^2 \left(\frac{t_U}{t_{contr}}\right)^2 \left[\frac{\rho_m (t)}{\rho_{m0}} \right]^{3n+1}
\left(\frac{\rho_c}{\rho_{m0}} \right)^{2n+2} > 1.
\label{spike-region}
\ee
The values of the densities $\rho_{m0}$ and $\rho_m (t)$ depend upon the objects under scrutiny. If we speak about
formation of galaxies or their clusters the following ratios can be expected: $\rho_{m0}/\rho_c = 1 - 10^3$ and 
$\rho_{m} (t) / \rho_{m0}  $ varying  in the range $1 - 10^5$.  Indeed the oscillations of  curvature in such systems are excited 
if their mass density started to rise with time. For large scale structures this process began when they decoupled from 
the overall Hubble flow, which mostly took place for redshift in the interval $z = 10 - 0$, 
and could result in creation of galaxies with
the energy density 5 orders of magnitude higher than the present day cosmological one.
If we consider formation of stellar or planetary type objects from the intergalactic gas with the initial density 
$10^{-24} $ g/cm$^3$, then $\rho_{m0}/\rho_c = 10^5$ and $\rho_m (t) / \rho_{m0} $ can vary in the range
$ 1 - 10^{24}$ or more. 

The analysis of ref.~\cite{Arbuzova:2012su,Arbuzova:2013ina} has been done under the assumption that the background 
space-time is nearly flat and so the background metric is almost Minkowsky. However, the large deviation of curvature from its GR value, found in these works, may invalidate the assumption of an approximately flat background and should be verified.
In what follows we consider a spherically symmetric bubble of matter, e.g. a gas cloud or some other astronomical object, which occupies a finite region of space of radius $r_m$, and study spherically symmetric solution of corresponding equations of motion, assuming that the metric has the Schwarzschild form:
\be
ds^2 =  A (r,t) dt^2 - B(r,t) dr^2 - r^2 (d\theta^2 + \sin^2 \theta\,d\phi^2) .
\label{ds2}
\ee

We assume that the metric is close to the flat one, i.e.
\be
A_1 = A - 1 \ll 1\,\,\,{\rm and} \,\,\, B_1 = B -1 \ll 1
\label{A1-B1}
\ee
and study if and when this assumption remains true for the solutions with very large values of $R$ found in our previous works~\cite{Arbuzova:2012su,Arbuzova:2013ina}. 

We construct the internal solution assuming that it consists of two terms: the Schwarzschild one and the oscillating part generated by the rising density. 
As it was found in our work \cite{Arbuzova:2013pta}, 
the metric functions inside the cloud are equal to:
\be
B (r, t) & =& 1 + \frac{2M(r,t)}{M_{Pl}^2r} \equiv  1+ B_1^{(Sch)}\, ,
\label{B-of-r-t}\\ 
A(r,t)  &=& 1 + \frac{R(t )\,r^2}{6} + A_1^{(Sch)} (r,t) \label{A-of-r-t}\, .
\ee
Here $M(r,t)$ is a mass of matter inside a radius $r$.

For the Schwarzschild part of the metric function $A(r,t)$ we have: 
\be
A_1^{(Sch)} (r,t) = \frac{r_g r^2}{2r_m^3} -\frac{3r_g}{2r_m}+
 \frac{ \pi \ddot\varrho_m}{3 M_{Pl}^2}\, ( r_m^2 - r^2)^2\, ,
\label{A-1-2}
\ee
where $r_g=2M/M_{Pl}^2$ is the usual Schwarzschild radius. 

As we noted, 
$R(t)$ is typically larger that the GR value: $|R_{GR}|=8 \pi \varrho_m /M_{Pl}^2$, so
the second term in eq. (\ref{A-of-r-t}),  $R(t)r^2/6$, 
gives the dominant contribution into $A_1$ at sufficiently large $r$. 
Indeed, $r^2 R(t) \sim r^2 y R_{GR}$ with $y > 1$, while the canonical Schwarzschild terms are of the order 
of $r_g/r_m \sim \varrho_m r_m^2/m_{Pl}^2 \sim r_m^2 R_{GR} $. 

As is already mentioned, the solution with large oscillating $R(t)$ was obtained~\cite{Arbuzova:2012su,Arbuzova:2013ina} 
under the assumption that the background metric weakly
deviates from the flat Minkowsky one. Though this is certainly true for the Schwarzschild part of the solution 
(\ref{A-1-2}), this may be questioned for the $r^2 R(t)/6$ - term. 
Evidently the flat background metric is not noticeably distorted if $ r ^2<  6/R(t)$.
If the initial energy density of the cloud is of the order of the cosmological energy density, i.e. $R_{GR}\sim 1/t_U^2$, 
then the metric would deviate from the Minkowsky one for clouds having radius $r_m > t_U/\sqrt y$, 
where the maximum value of $y$ is given by eq. (\ref{y-of-t}). For systems where very large values of $y$ 
are reached, the flat space approximation may be broken already for non-interestingly small $r$. However, at the
stage of rising $R(t)$ when $y>1$ but not huge, the flat space approximation would be valid over all the
volume of the collapsing cloud. 
For large objects or large $y$, such that $R r^2 /6 \sim 1$, the approximation of flat background metric 
becomes inapplicable and one has to solve the exact non-linear equations;
this situation will be studied elsewhere. If $A_1$ becomes comparable with unity, the evolution of $R(t)$ may 
significantly differ from that found in~\cite{Arbuzova:2012su,Arbuzova:2013ina}, but it seems evident that once a large $y>1$ is reached,
it would remain larger than unity despite a possible back-reaction of the non-flat metric.

In the lowest order in the gravitational interaction the motion (the geodesic equation in metric (\ref{ds2}))
of a non-relativistic  test particle is governed by the equation:
\be
\ddot r = - \frac{A'}{2} = -\frac{1}{2}\left[ \frac{R(t) r}{3} + \frac{r_g r}{r_m^3} \right],
\label{ddot-r}
\ee
where $A$ is given by eq. (\ref{A-of-r-t}). Since $R(t)$ is always negative and large, the modifications of GR considered here lead to 
anti-gravity inside a cloud with energy density exceeding the cosmological one. Gravitational repulsion dominates over the usual attraction if
\be
\frac{|R|r_m^3}{3r_g} = \frac{|R|r_m^3M_{Pl}^2}{6 M} = \frac{|R|M_{Pl}^2}{8\pi \varrho}  
\equiv y > 1\,,
\ee
so basically whenever  oscillations of  $R$ start rising, regardless of the initial value of $\varrho$ and to some extent of the 
specific $F(R)$ considered. Therefore, this is most likely a more fundamental statement, applicable to essentially all $F(R)$ models 
producing oscillations of $R$  with the amplitude larger than the GR value.

 So, in modified gravity and in systems with rising energy density, the curvature scalar would typically exceed the GR value $R_{GR}$, i.e. $y>1$, and thus the gravitational repulsion would dominate over the usual Schwarzcshild attraction. The back-reaction
of this repulsion would slow down the contraction but evidently do not stop it.  Moreover, the repulsion could overtake the 
contraction at sufficiently large radius. As a result shell type structures could be formed.
Sufficiently 
large primordial clouds would not shrink down to smaller and smaller bodies with more or less uniform density 
but could form thin shells empty (or almost empty) inside, except possibly for some central mass.  
Hence the gravitational repulsion found here might be responsible for the formation of cosmic voids but the 
lengthy analysis of realistic scenarios is outside the framework of the presented letter. 

One more comment may be in  order here. In the standard GR the Jebsen-Birkhoff theorem holds (see 
e.g. the book~\cite{Hawking:1973uf}), which essentially says that any finite body with positive definite energy density
always has  an attractive gravitational action. Our result is in clear contradiction with this statement.  The  Jebsen-Birkhoff theorem
in the case of modified gravity is discussed in detail in the review~\cite{Capozziello:2011et}, see also ref.~\cite{Capozziello:2007ms}.
It is shown that "in a space-time with constant scalar curvature, any spherically symmetric background is necessarily
static or, the Jebsen-Birkhoff  theorem holds for $f(R)$-gravity with constant curvature", while it breaks if curvature is not
constant, as is the case considered in our work.

\section{Gravitational instability in quickly oscillating background \label{sec-Jeans}} 

In this section we consider evolution of density perturbations in modified gravity. Since a Lagrandian is a non-linear function of curvature, $R$, 
the equation of motion becomes of higher (4th) order and the evolution of perturbations may differ significantly from that in General Relativity. 

In cosmology this problem has been consider for different forms $F(R)$ in Refs. \cite{Zhang:2005vt,Song:2006ej,Tsujikawa:2007gd,delaCruzDombriz:2008cp,Ananda:2008tx,Ananda:2008gs,Motohashi:2009qn,Matsumoto:2014wca}.
A analysis of the Jeans instability for the stellar-like objects in modified gravity was performed in works 
~\cite{Capozziello:2011nr,Capozziello:2011gm,Eingorn:2014laa}. 
In our works \cite{Arbuzova:2014bha,Arbuzova:2015uga} we study the associated instabilities not only in quasi stationary background, but also in a 
 background of quickly oscillated curvature.  The time evolution of first-order perturbations is governed by a fourth-order 
differential equation instead of the usual second order one, therefore new types of unstable solutions are induced.
 
 There appear not only the usual Jeans solution with a slightly reduced length scale, but also parametric resonance amplification of density perturbations, and an amplification of the perturbations due to the
 "antifriction" behaviour of the coefficients of odd derivatives in the equation.

We assume that the background metric weakly deviates from the Minkowski one, while derivatives of the metric may be far from their GR values. In particular, $R$ may be very much different from $R_{GR} = - 8 \pi T_\mu^\mu/M_{Pl}^2 \equiv -\tilde T$, where $ T_\mu^\mu$ is the trace of the energy-momentum tensor of matter. 

We consider the case $|R_c| \ll |R| \ll m^2$, which is realized in astronomical systems with the energy density grossly exceeding the cosmological one. 
In this regime the modified equation of motion can be approximated as 
\be
G_{\mu\nu} + \frac{1}{3\omega^2}(D_{\mu}D_{\nu} - g_{\mu\nu}D^2)R = \tilde T_{\mu \nu}\, ,
\label{EoM-F(R)}
\ee
where $G_{\mu\nu} = R_{\mu\nu} - g_{\mu\nu} R/2$ is the Einstein tensor and $\omega ^{-2} = - 3 F''_{RR}$. 

Once written in this form, the equation is largely independent of the specific model considered except of course of the value of $\omega$.
 
As previously, we consider a spherically symmetric cloud of matter with initially constant energy density inside the limit radius $r=r_m$. We choose Schwarzschild-like isotropic coordinates in which the metric takes the form:
\begin{eqnarray}
ds^2=Adt^2 - B\,\delta_{ij}\,dx^i dx^j\,, 
\label{ds-2}
\end{eqnarray}
where the functions $A$ and $B$ depend upon $r$ and $t$. 

As usually the metric and the curvature tensors are expanded around their background values at first order in infinitesimal perturbations, i.e.
\begin{eqnarray}
A  = A_b + \delta A\,, \ \ 
B  = B_b + \delta B\,,\ \ \
R  = R_b + \delta R\,,
\end{eqnarray}
where $A_b$ and $R_b$ are quickly oscillating functions of time with possibly large amplitude ("spikes"), discussed in previous sections. 

We assume 
that ${ \omega = const}$ and  study the development of instabilities described by 
{ the fourth order differential equation, which governs evolution of perturbations in this model.} 
In this case, the evolution of instabilities is quite different from { the standard situation} described by { the second order equation of { GR}. }
We will not dwell on a particular choice of the ${F(R)}$-function,
but assume that the high frequency oscillations of $ R$
are a generic phenomenon in such models.   

Using technique described in papers~\cite{Arbuzova:2014bha,Arbuzova:2015uga} we derive the 4th order equation for the single function $\delta B$.
We introduce the dimensionless time and define:
\be
\label{definitions_abc}
\tau=\omega t\,, \ \ \delta B \equiv z\,,\ \
 R_b = - \tilde \rho_b y \,, \ \ 
 a = \frac{\tilde\rho_b}{ k^2}\,,\ \
 b =\frac{k^2}{\omega^2}\,,\ \
 c = c_s^2\,.
\ee
With these quantities equation takes a form convenient for the analysis:
\be
z'''' + \alpha y' z''' + (\Omega^2 + 2\alpha y'')z'' + \alpha(y''' + bc y')z' + \mu z = 0\,,
\label{z-4order-abc}
\ee
where
\be
 \label{definitions_Omega_mu}
\alpha &= &\frac{a}{2}\left(1+\frac{2b}{3}\right)\,, \ \ 
\Omega^2  = 1 - \frac{a}{2}\left(1+\frac{8b}{3}\right) + b(1+c)\,,
\label{omega} \\
\mu & = &b \left[ c(1+b) - \frac{a}{2}\left(1 + \frac{4b}{3}\right) - 2ac\left(1+\frac{2b}{3}\right) \right].
\label{mu}
\ee

The 4th order equation,
governing evolution of metric perturbations, demonstrates very rich pattern of different types of instabilities.
If $y(\tau)$ is non-negligible, quite interesting new effects can show up. The function $y(\tau)$ entering the 4th order dimensionless equation~(\ref{z-4order-abc}) is an oscillating function of "time" $\tau$. It can induce an analogue of the parametric resonance instability resulting in a very fast rise of perturbations at a certain set of frequencies. Another new effect can be called "antifriction". It appears at sufficiently large amplitudes of oscillations of $y(\tau)$ such that the coefficients in front of the odd derivative terms
become periodically negative. This phenomenon leads to an explosive rise of $z$ in a wide range of frequencies. 

Both effects do not exist in the standard General Relativity and, if discovered, would be a proof of modified gravity. 
On the contrary, the non-observation of these effects would allow to put stringent restrictions on  the parameters of $F(R)$-theories.

\begin{figure}
\centering
 \includegraphics[width=.45\textwidth]{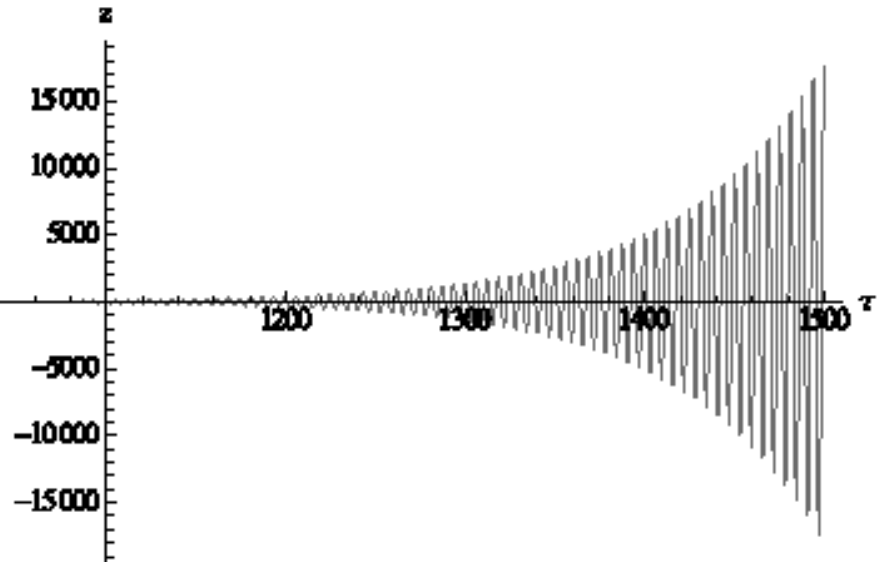}
 \includegraphics[width=.45\textwidth]{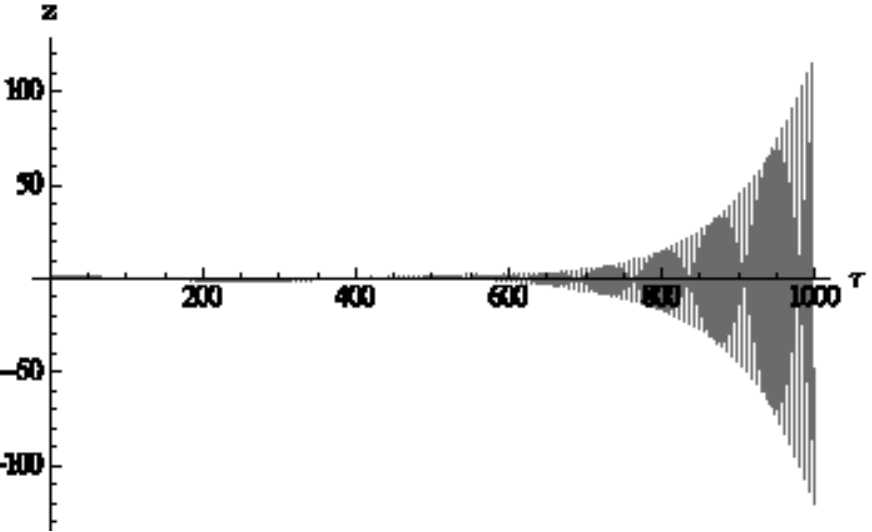}
\caption{{\it Left panel}: Parametric resonance excitation of $z(\tau)$ for harmonic curvature oscillations with $y_0=5$ and $\Omega_1/\Omega=2$.
{\it Right panel}: Parametric resonance excitation of $z(\tau)$ for spike-like curvature oscillations with $y_0=30$ and $\Omega_2/\Omega=0.5$.}
\label{fig:par-res}
\end{figure}

 We consider two possible forms of periodic function $y(\tau)$: 
 \begin{enumerate}
\item purely harmonic ones: 
\be
y_{harm}(\tau)= y_{eq}(\tau) + y_0\cos(\Omega_1 \tau + \theta)\,,
\label{y-harm}
\ee
where $\Omega_1 = \omega_1/\omega$ is the dimensionless frequency, $y_0$ is the amplitude of oscillations and $\theta$ is a constant phase,
and $y_{eq}$ is the equilibrium point of the potential around which the curvature oscillates.
\item spiky solutions found in Ref.~\cite{Arbuzova:2012su}, which we approximate as 
\be
y_{sp}(\tau)=\frac{y_0\,d^2}{d^2 + \sin^2(\Omega_2 \tau + \theta)}\,,
\label{y-sp}
\ee
where $d\ll 1$, so that we have narrow peaks with a large separation between them. 
\end{enumerate}
We solved Eq.~(\ref{z-4order-abc}) numerically for different values of $y_0$ and $\Omega_{1,2}$. The parametric resonance excitation for harmonic oscillations is observed at the expected frequency $\Omega_1 = 2\Omega$, see Fig.~\ref{fig:par-res} (left panel). Here and in what follows the parameters of the medium and the wave number of the fluctuations were taken according to $a=b=0.01$ and $c=c_s^2 =0.02$. 
Fig.~\ref{fig:par-res} (right panel) clearly shows parametric resonance in the case of spike-like oscillations at $\Omega_2=\Omega/2\simeq 0.5013$ which corresponds to the second mode of the resonance, see Eq.~(\ref{omega}).
 The parametric resonance is rather sensitive to variations of $y_0$ and of course of $\Omega_{1,2}$. In fact, the exponential growth is slower as we depart from the resonance frequency.

\begin{figure}[ht]
\centering
 \includegraphics[width=2.5in]{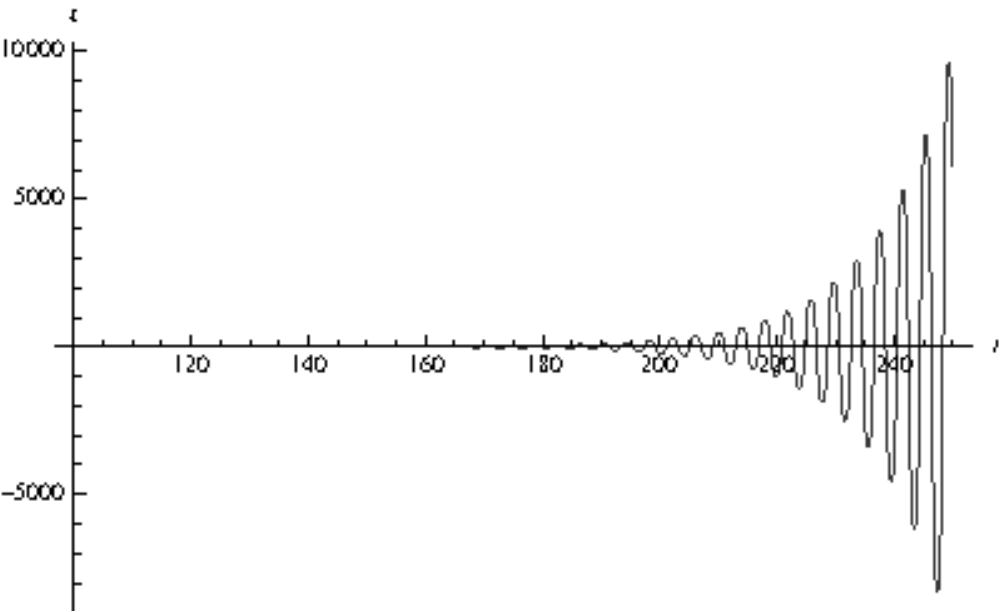}
 \includegraphics[width=2.5in]{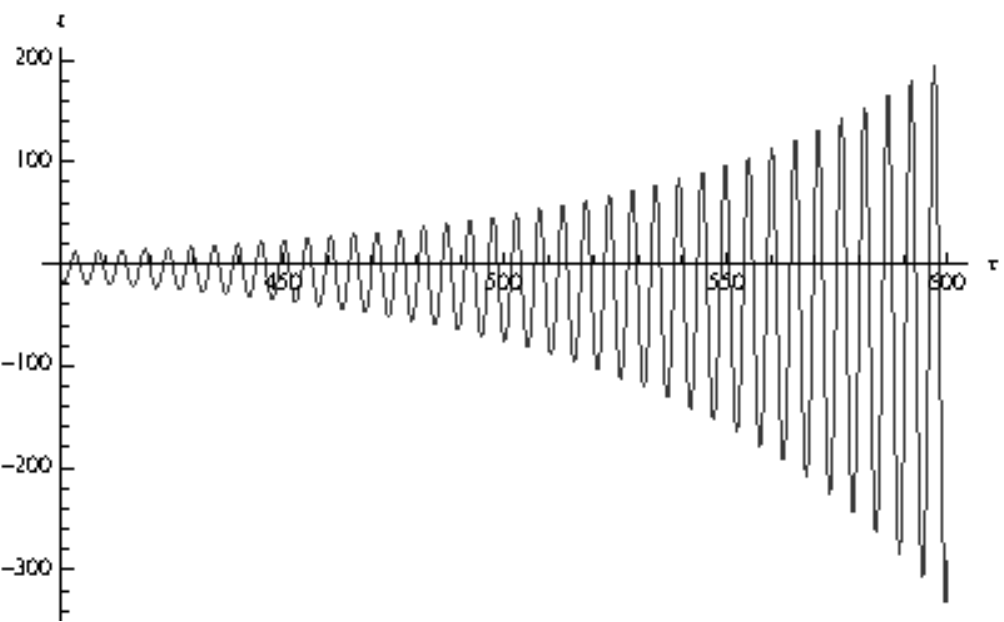}
 \caption{\textit{Left panel}: Antifriction effect in evolution of $z(\tau)$ for harmonic curvature oscillations with $y_0=169$, $\Omega_1/\Omega=3.2$.
\textit{Right panel}: Antifriction effect in evolution of $z(\tau)$ for spike-type curvature oscillations with $y_0=400$, $\Omega_2/\Omega=0.6$. }
\label{fig:antifric}
\end{figure}

 The antifriction amplification is observed at the frequencies away from the resonance values, if $y_0$ exceeds a threshold value, $y_{th}$. The farther away the frequency is from the resonance, the larger is the threshold. 
For example for harmonic oscillations with $\Omega_1/\Omega = 3.2$ the threshold value is $y_{th} = 169$, while for $\Omega_1/\Omega = 4.4$ the threshold is $y_{th} = 267$. 
The evolution of $z(\tau)$ in the first case is depicted in Fig.~\ref{fig:antifric} (left panel).

In the right panel of Fig.~\ref{fig:antifric} we present the evolution of $z(\tau)$ for spike-like oscillations with out-of-resonance frequencies $\Omega_2/\Omega = 0.6$. Similarly to the case of harmonic $y(\tau)$, the farther away the frequency is from the resonant one, approximately equal to $0.5$, the larger the threshold value of $y_0$ necessary for generating an unstable solution. 

Since the physically interesting quantity is the magnitude of density perturbations, we present the relative density contrast,
${\delta \rho/\rho_b}$, expressed through perturbations of metric ${z \equiv \delta B}$:
\be 
\frac{ \delta \rho}{\rho_b}=z}\left[{\frac{1+b}{a(1+2b/3)} - 2}\right] {+ \frac{1}{2}\,z'\,y' + \frac{z''}{a (1+2b/3)}\,.
\label{delta-rho-f}
\ee  

Exponentially growing solutions for $z$ will lead to an equivalent behaviour for the density perturbation $\delta\rho$, as shows in Fig.~\ref{fig:delta-rho} in the case of parametric resonance induced by the spiky solution with moderately large amplitude $y_0 = 30$.

\begin{figure}[ht]
\centering
 \includegraphics[width=.48\textwidth]{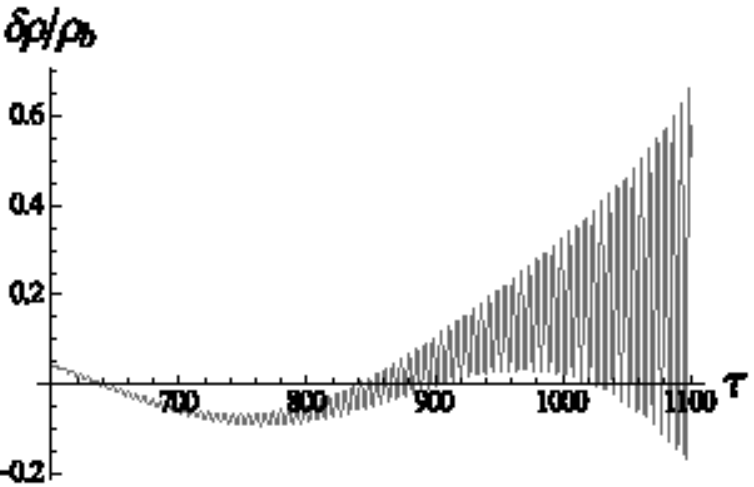}
\caption{Evolution of ${\delta \rho/\rho_b}$ in parametric resonance region induced by the spike-type action with ${y_0=30}$, 
${\Omega_2/\Omega = 0.5}$, 
The initial value of metric perturbation is taken as 
${\delta B(0) \equiv z(0) = 10^{-3}}$.}
\label{fig:delta-rho}
\end{figure}

\section{Modified gravity in Early Universe \label{sec-DM}}

In previous sections we considered the models based on $F(R)$-function in the form (\ref{eq:model}) with $R^2$-term included to prevent from the singular behaviour both in the past and in the future:
\be \nonumber
F(R) = -\lambda R_c\left[1-\left(1+\frac{R^2}{R_c^2}\right)^{-n}\right]-\frac{R^2}{6m^2}\,,
\ee
 However, in very early universe $R^2$-term, introduced for the elimination of the curvature 
singularity is dominant, while the first term is not essential.  

In 1979 V.Ts. Gurovich and A.A. Starobinsky~\cite{Gurovich:1979xg} suggested to take $F(R)=-R^2/(6m^2)$ for elimination of 
cosmological singularity.  In the subsequent paper by Starobinsky~\cite{Starobinsky:1980te} it was found that the 
addition of the $R^2$-term leads to inflationary cosmology. 

In our work \cite{Arbuzova:2018ydn} it was shown that $R^2$-term creates  considerably deviation from the Friedmann cosmology in post-inflationary epoch. Gravitational particle production 
by the oscillating curvature, $R(t)$, led to a graceful exit from inflation, but the cosmological evolution in the early universe was 
drastically different from the standard one till the universe age reached the value of the order of the inverse decay rate of the
oscillating $R(t)$. This deviation from the standard cosmology leads to reconsideration of  the 
of primordial black holes formation, modifies high temperature baryogenesis, opens the window for heavy Lightest Supersymmetric Particles (LSPs) to be the cosmological Dark Matter (DM).

Since the first indications to existence of dark matter by Kapteyn \cite{Kapteyn:1922zz}, Oort \cite{Oort32} and 
  Zwicky~\cite{Zwicky:1933gu} in 1933 and later confirmation in 1970s \cite{Rubin:1970zza,Einasto-Nature-250,Ostriker:1974lna} 
  many theoretical models have been proposed to describe this elusive form of matter. 
  
An accepted property of  the DM particles is that they are electrically neutral\footnote{It is nevertheless possible that dark matter particles
have a tiny electric charge or even a "normal"  charge but very high mass, so the Thomson scattering is strongly suppressed. }
 since they don't scatter light, hence the 
name \textit{Dark} matter. Otherwise their properties are practically unknown.
This opened possibilities for innumerable particles to be DM candidates.  
  
A natural, and formerly very popular, candidate for dark matter particle is the lightest supersymmetric  particle  which should be stable if the so-called R-parity is conserved. The latest reviews on SUSY dark matter, and not only, can be found in 
Refs.~\cite{Catena:2013pka,Gelmini:2015zpa,Lisanti:2016jxe,Slatyer:2017sev,Cline:2018fuq}.
 
An extensive search for the low energy supersymmetry performed at LHC led to negative results. Thus, if supersymmetry exists, its characteristic energy scale should be, roughly speaking, higher than 10 TeV. The cosmological energy density of LSPs is proportional to their mass squared, 
$\rho_{LSP} \sim m_{LSP}^2$, and for $m_{LSP} \sim 1$~Tev $\rho_{LSP}$ is of the order of the observed energy density of the universe.
Correspondingly for larger masses such particles would overclose  the universe. This  unfortunate 
circumstance excludes LSPs as dark matter particles in the conventional cosmology. 

There are several attempts in the literature to save  supersymmetric dark matter by modifying the cosmological
scenarios of LSP production in such a way that the relic density of heavy LSP would be significantly suppressed.
For example in the paper~\cite{kane-non-therm} a detailed study of non-thermal production of heavy relics is performed.
Recently in Ref.~\cite{Drees:2018dsj} a specific scenario has been studied, which is based on the assumption
that after the freezing of LSP the universe was matter dominated and this epoch
transformed into the radiation dominated stage with low reheating temperature.  Similar idea was discussed earlier in 
paper~\cite{DNN}, where it was assumed that at some early stage the universe might be dominated by primordial black holes
which created the necessary amount of entropy to dilute the heavy particle relics.

The $R^2$ inflation was generalized to supergravity in the series of 
papers~\cite{Ketov:2010qz,sg2,Ketov:2018uel} and references therein.
In particular in Ref.~\cite{sg2} a scenario with superheavy gravitino, which 
may be a viable candidate for  dark matter particle, was considered. 

In our paper \cite{Arbuzova:2018apk} 
we have shown that in $(R+R^2)$-gravity the energy density of LSPs may be much lower and they are viable candidates for the dark matter 
particles, if their mass is about 1000 TeV.  The mechanism  considered there is different from that proposed in  papers cited above.

\section{Conclusions}
 
\begin{enumerate}
\item
{
A general feature of ${F(R)}$ modified gravity is high frequency and large amplitude oscillations 
of curvature and metric in contracting matter systems. }
\item 
{
These oscillations lead to the production of elementary particles, which may be observable 
in the spectra of energetic cosmic rays. 
\item
In the background of this oscillating solution, gravitational repulsion between objects of finite size
is possible. Such a repulsion might be responsible for the creation of the observed cosmic voids.
\item
The process of structure formation in modified gravity could be both amplified and suppressed in  
comparison with  General Relativity.
\item
The fourth order equations,
governing evolution of the density perturbations, demonstrate very rich pattern of different types of instabilities:
an analogue of parametric resonance} and the antifriction instability.   
\item
The early universe evolution  in $R^2$-theory significantly deviates from that
in the Friedmann cosmology. 
\item
The supersymmetric particles with $m_{LSP} \gtrsim 1000$ TeV are viable candidates for the constituents of Dark Matter.
\end{enumerate}

\bigskip
\bigskip
\noindent {\bf DISCUSSION}

\bigskip
\noindent {\bf DANIELE FARGION:} Does this new $R+R^2$ term model produce 
some donations to the geodesic trajectories around black hole (BH) like our galactic centre or
host M87 BH shadows?

\bigskip
\noindent {\bf ELENA ARBUZOVA:} The effects of any known gravity modification at these 
spacial scales are practically unnoticeable.

\end{document}